\documentclass[twocolumn,aps,nofootinbib,prl,showpacs,floatfix]{revtex4}
\usepackage{graphicx, epsfig, array}

\def\be{\begin{equation}}
\def\ee{\end{equation}}
\def\bea{\begin{eqnarray}}
\def\eea{\end{eqnarray}}

\def\reff@jnl#1{{\rm#1\/}}

\def\aj{\reff@jnl{AJ}}                  
\def\araa{\reff@jnl{ARA\&A}}            
\def\apj{\reff@jnl{ApJ}}                        
\def\apjl{\reff@jnl{ApJ}}               
\def\apjs{\reff@jnl{ApJS}}              
\def\ao{\reff@jnl{Appl.Optics}}         
\def\apss{\reff@jnl{Ap\&SS}}            
\def\aap{\reff@jnl{A\&A}}               
\def\aapr{\reff@jnl{A\&A~Rev.}}         
\def\aaps{\reff@jnl{A\&AS}}             
\def\azh{\reff@jnl{AZh}}                        
\def\baas{\reff@jnl{BAAS}}              
\def\jrasc{\reff@jnl{JRASC}}            
\def\memras{\reff@jnl{MmRAS}}           
\def\mnras{\reff@jnl{MNRAS}}            
\def\pra{\reff@jnl{Phys.Rev.A}}         
\def\prb{\reff@jnl{Phys.Rev.B}}         
\def\prc{\reff@jnl{Phys.Rev.C}}         
\def\prd{\reff@jnl{Phys.Rev.D}}         
\def\prl{\reff@jnl{Phys.Rev.Lett}}      
\def\pasp{\reff@jnl{PASP}}              
\def\pasj{\reff@jnl{PASJ}}              
\def\qjras{\reff@jnl{QJRAS}}            
\def\skytel{\reff@jnl{S\&T}}            
\def\solphys{\reff@jnl{Solar~Phys.}}    
\def\sovast{\reff@jnl{Soviet~Ast.}}     
\def\ssr{\reff@jnl{Space~Sci.Rev.}}     
\def\zap{\reff@jnl{ZAp}}                        
\def\nat{\reff@jnl{Nature}}             

\def\fun#1#2{\lower3.6pt\vbox{\baselineskip0pt\lineskip.9pt
        \ialign{$\mathsurround=0pt#1\hfill##\hfil$\crcr#2\crcr\sim\crcr}}}
\renewcommand\({\left(}
\renewcommand\){\right)}

\newcommand\eV{\rm eV}

\newcommand\lsim{\mathrel{\rlap{\lower4pt\hbox{\hskip1pt$\sim$}}
    \raise1pt\hbox{$<$}}}
\newcommand\gsim{\mathrel{\rlap{\lower4pt\hbox{\hskip1pt$\sim$}}
    \raise1pt\hbox{$>$}}}

\def\dslash{\not{\hbox{\kern-2pt $\partial$}}}
\def\Dslash{\not{\hbox{\kern-4pt $D$}}}
\def\Oslash{\not{\hbox{\kern-4pt $O$}}}
\def\Qslash{\not{\hbox{\kern-4pt $Q$}}}
\def\pslash{\not{\hbox{\kern-2.3pt $p$}}}
\def\kslash{\not{\hbox{\kern-2.3pt $k$}}}
\def\qslash{\not{\hbox{\kern-2.3pt $q$}}}

 \newtoks\slashfraction
 \slashfraction={.13}
 \def\slash#1{\setbox0\hbox{$ #1 $}
 \setbox0\hbox to \the\slashfraction\wd0{\hss \box0}/\box0 }

\begin{document}

\setlength{\unitlength}{1mm}

\title{Detecting neutrino mass difference with cosmology}

\author{An\v{z}e Slosar}
\affiliation{Faculty of Mathematics and Physics, University of Ljubljana, Slovenia}

\date{\today}%

\begin{abstract}
  
  Cosmological parameter estimation exercises usually make the
  approximation that the three standard neutrinos have degenerate
  mass, which is at odds with recent terrestrial measurements of the
  difference in the square of neutrino masses. In this paper we
  examine whether the use of this approximation is justified for the
  cosmic microwave background (CMB) spectrum, matter power spectrum
  and the CMB lensing potential power spectrum. We find that, assuming
  $\delta m^2_{23} \sim 2.5 \times 10^{-3}$eV$^2$ in agreement with
  recent Earth based measurements of atmospheric neutrino
  oscillations, the correction due to non-degeneracy is of the order of
  precision of present numerical codes and undetectable for the
  foreseeable future for the CMB and matter power spectra. An
  ambitious experiment that could reconstruct the lensing potential
  power spectrum to the cosmic variance limit up to $\ell \sim 1000$
  will have to take the effect into account in order to avoid biases.
  The degeneracies with other parameters, however, will make the
  detection of the neutrino mass difference impossible.  We also show
  that relaxing the bound on the neutrino mass difference will also
  increase the error-bar on the sum of neutrino masses by a factor of
  up to a few. For exotic models with significantly non-degenerate
  neutrinos the corrections due to non-degeneracy could become
  important for all the cosmological probes discussed here.

\end{abstract}
\bigskip

\maketitle

\section{Introduction}

Standard cosmological measurements offer an excellent probe of
neutrino physics \cite{1996ApJ...467...10D}. In contrast to Earth
based measurements of neutrino oscillations that measure the
difference in the square of neutrino mass eigen states, the cosmology
is sensitive to the absolute mass of neutrinos. At the moment,
cosmology is the only viable alternative to the beta-decay experiments
\cite{Farzan:2001cj} in this field and exceeds it in accuracy.

The most natural and accurate way to measure neutrino masses with
cosmology is via measurements of the power spectrum of fluctuations in
the cosmic microwave background (CMB) and the power spectrum of matter
fluctuations, which is one of the basic products of the galaxy
redshift surveys. In the future, the non-Gaussianity introduced by
the lensing of the CMB fluctuations by the intervening structures between the last
scattering surface and us will prove to be an important tool to
constrain the power spectrum of matter fluctuations and consequently
neutrino masses \cite{Lesgourgues:2005yv, 2006astro.ph..1594L,Hannestad:2006zg}.

Many of the recent parameter estimation papers have put bounds on the
sum of the neutrino masses
\citep{2003ApJ...583....1B,2003JCAP...05..004H,2003astro.ph..6386A,2005astro.ph..7503M,2004PhRvD..69j3501T,
  2005PhRvD..71j3515S,2004PhRvD..70k3003F,Barger:2003vs,Crotty:2004gm,Hannestad:2004bu,Hannestad:2005gj,Tegmark:2005cy},
with the upper bound on the sum of neutrino masses of the order
$0.5$eV-$1$eV. The three neutrino species were assumed to be
degenerate in mass.

However, Earth based measurements of neutrino oscillations suggest not
only that neutrinos have mass, but also that these masses are not
equal \cite{Fogli:2005cq, Maltoni:2004ei}.  Neutrino that is produced
in a flavour eigenstate, which is some linear superposition of mass
eigenstates, can latter be measured as being of a different flavour,
due to different mass eigenstates acquiring a different phase during
propagation.  In particular, oscillations of atmospheric neutrinos
suggest a squared mass difference of $|\Delta m^2_{23}| =
|m_3^2-m_2^2| = \sim 2.5 \times 10^{-3} \eV^2$
\cite{2001PhRvL..86.5656F,2004PhRvL..93j1801A}, while solar neutrino
observations, together with results from the KamLAND reactor neutrino
experiment, point towards $\Delta m^2_{12} = m_2^2-m_1^2 = \sim 5
\times 10^{-5} \eV^2$ \cite{2001PhRvL..87g1301A,2002PhRvC..66c5802B}.
Note that in the former case, the sign of the difference is not know,
while in the latter it is. Given these constraints, it is possible
construct two hierarchies of masses. Assuming the lightest neutrino to
be of a negligible mass, the neutrino masses can be either in a
so-called normal hierarchy with masses around $\sim0, \sim\sqrt{\Delta
  m^2_{12}}, \sim\sqrt{\Delta m^2_{23}}$ or in an inverted hierarchy,
in which case masses are $\sim0, \sim\sqrt{|\Delta m^2_{23}|},
\sim\sqrt{|\Delta m^2_{23}|}$. This means that it is possible to rule
out the inverted hierarchy simply by measuring the sum of neutrino
masses and excluding the $\sum m>2\sqrt{\Delta m^2_{23}}$ region.

Another way of distinguishing between the two hierarchies is to try to
measure the neutrino mass difference directly with cosmology. The
above results imply that the three neutrino families are
non-degenerate at the level of $\sim 5$\% if the sum of neutrino
masses is 0.5eV.  Decreasing the sum to 0.2eV, the two neutrinos are
non-degenerate at the level of 25\%. Hence, it is timely to
investigate what are the biases introduced by assuming degenerate
neutrino masses in cosmological probes of neutrinos and this is
exactly what this paper is set to do.

Significant amounts of work in this direction was done before in
\cite{Lesgourgues:2004ps}. This paper focuses on the ability of future
Large Scale Structure (LSS) and CMB experiments to constrain neutrino
masses by performing the Fisher matrix analysis. They were the first
to perform the numerical integration of the CMB and LSS power spectra
by independently integrating more than one neutrino species. We extend
their work in several aspects. Firstly, we also include the lensing
potential reconstruction in the set of datasets used to constrain
neutrino masses. This has been studied in antoher recent paper
\cite{Lesgourgues:2005yv}, which shows that it would be impossible to
distinguish between normal and inverted hierarchies using even a very
optimistic future experiment. Nevertheless, the authors of that paper
do not use the multi-neutrino code and do not discuss degeneracy
assumption further. Secondly, rather than expanding around a fiducial
model, we study the general parameter space (although we also provide
Fisher matrix analysis in order to check the effect of degeneracies).
Finally, we provide a confirmation of their results by an independent
implementation of a linear code with more than one neutrino species.
An analysis of future sensitivities of high redshift galaxy surveys
and CMB data to measure neutrino masses and number of neutrino species
can be found in \cite{Takada:2005si}.

Assuming that the standard physics of the early universe applies, the
energy density and mass of a given neutrino species are related through
\begin{equation}
  w_{\nu} = \frac{m_\nu}{94.2\rm eV}.
\end{equation}
Interactions in the early universe before neutrino decoupling ensure
decoherence  and so flavour physics does not enter cosmology.
On the other hand, changing energy density (or mass) of a given neutrino species,
affects the CMB and matter power spectra in two ways. First,
it changes the redshift of the matter-radiation equality, thus
affecting the position and height of the peaks in the CMB power
spectrum and the maximum in the matter power spectrum. Second, it
damps the power spectrum on small scales, because relativistic
neutrinos in the early universe behave as radiation and effectively
stream away from over-dense regions. The free streaming wave-vector is
given by
\begin{equation}
  k_{\rm fs} \sim 0.01 \frac{m_\nu}{1 {\rm eV}}{\rm Mpc}
\end{equation}
and the  matter power spectrum is damped in  scales $k>k_{\rm fs}$.
For neutrinos of a small mass the second effect is considerably
more important.

In general, the neutrinos become non-relativistic at a redshift of
\cite{peima}

\begin{equation}
  z_{\rm nr} \sim \frac{m_\nu c^2}{3k_{\rm B} T_\nu} \sim 2 \times
  10^3 \left(\frac{m_\nu}{1{\rm eV}} \right) 
\end{equation}

This means that neutrinos lighter that about $0.5$eV will become
non-relativistic after the recombination and thus have a very small
effect on the CMB fluctuations power spectrum. The neutrino mass can,
however, be inferred from the gravitational lensing of the CMB
fluctuations.  The quantity of interest here is the projected
gravitational potential (see e.g. \cite{2003PhRvL..91x1301K})

\begin{equation}
  \phi(\mathbf{n}) = -2 \int_{0}^{r_\star} \Psi(\mathbf{n}r) \frac{r_\star-r}{rr_\star}{\rm d} r, 
\end{equation}
where $r$ are conformal distances, $r_\star$ is the conformal distance
to the surface of the last scattering and the integration is performed
along our past light cone. The spherical power spectrum of $\phi$,
$C_\ell^{\phi\phi}$ can be recovered using a variety of methods
\cite{2003PhRvD..67d3001H,2003PhRvD..68h3002H,1999PhRvL..82.2636S,2001ApJ...557L..79H,2000PhRvD..62d3007H}
and essentially contains information similar to that of the matter
power spectrum.

\begin{figure}
  \centering

  \includegraphics[angle=-90, width=\linewidth]{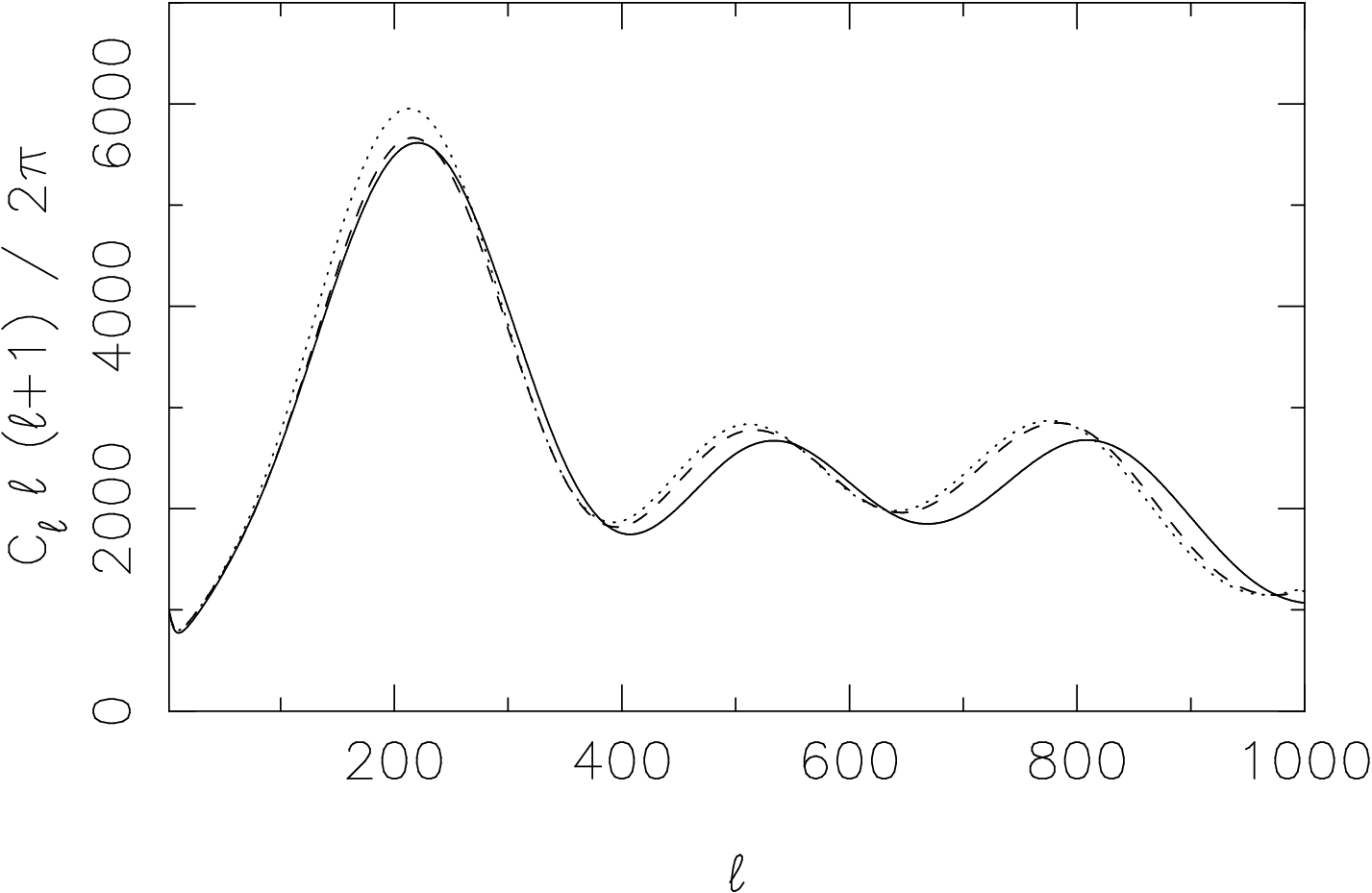}  
  \includegraphics[angle=-90, width=\linewidth]{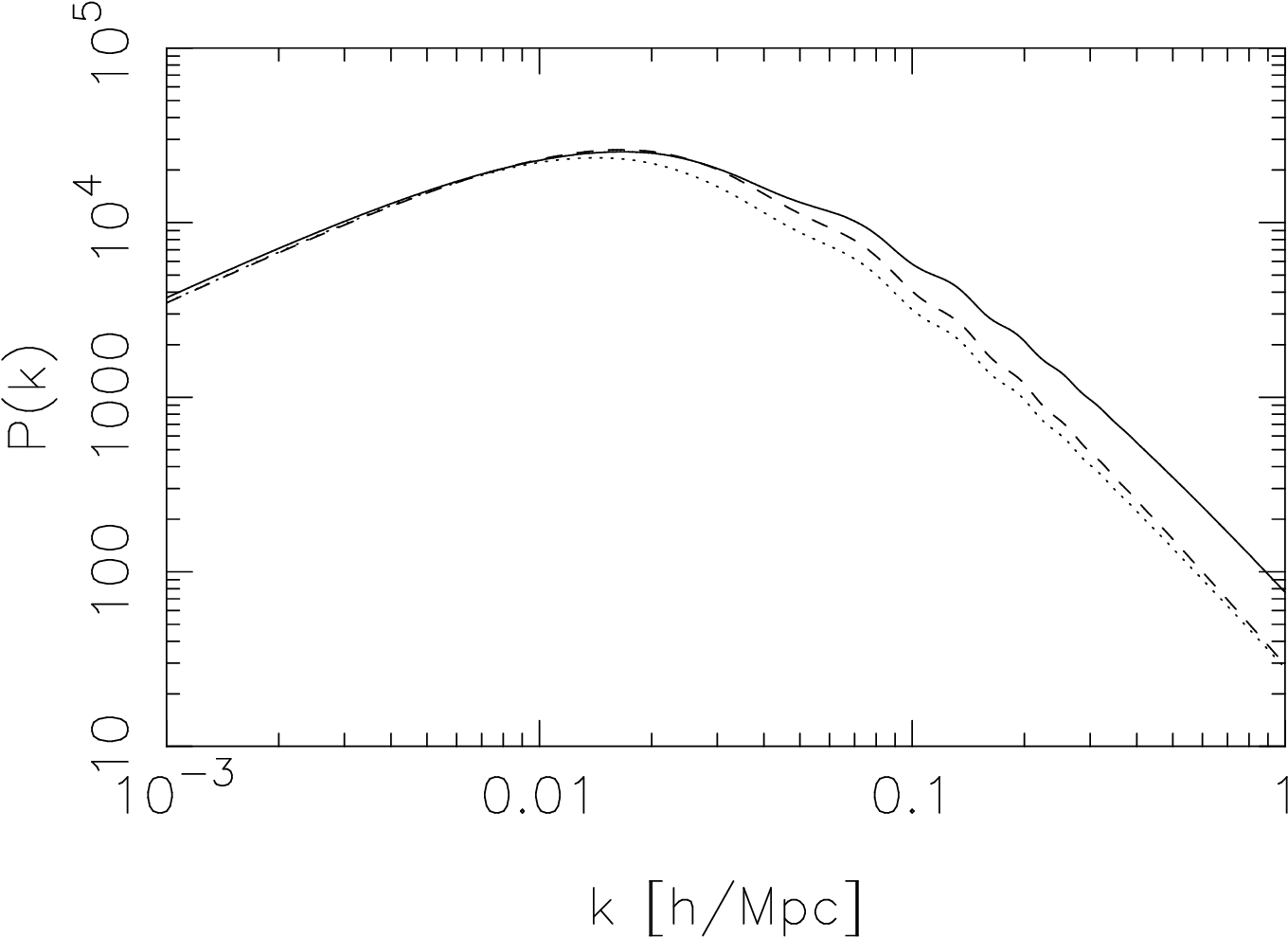} 
  \includegraphics[angle=-90, width=\linewidth]{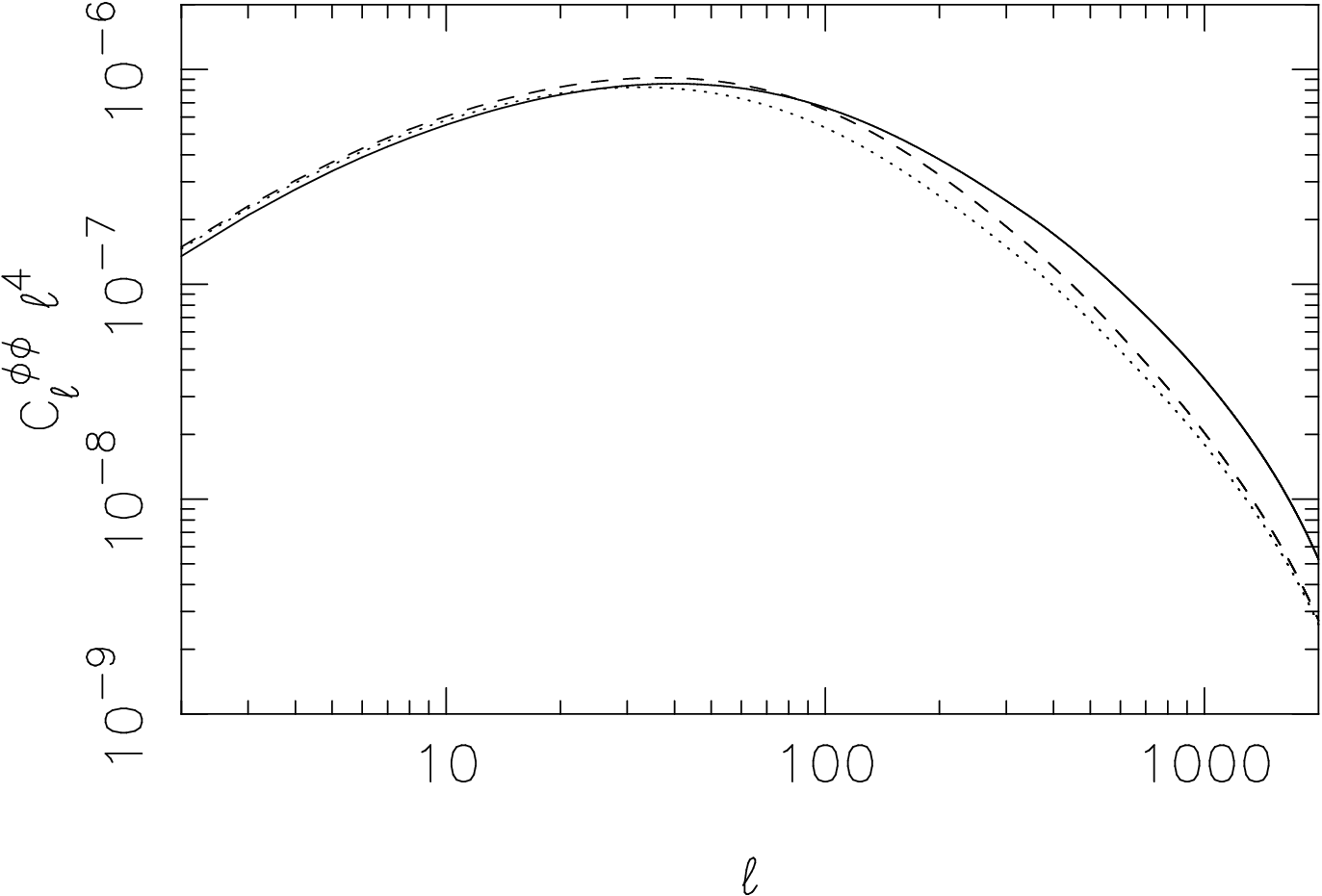}  

  \caption{The lensed CMB power spectrum (top), the matter power
  spectrum (middle) and the lensing potential power spectrum
  (bottom). Solid line correspond to the standard $\Lambda$CDM
  model. Other two models plotted have $\sum m_i=2$eV and $\alpha=1/3$
  (dotted) or $\alpha=1$ (dashed). See text for discussion.}
  \label{fig:big}
\end{figure}

\section{Non-degenerate neutrinos}

If one relaxes the degeneracy assumption, the neutrinos of different
masses become non-relativistic at different times and have different
free-streaming lengths, resulting in small corrections to the various
power spectra discussed above. Since the difference in the squares
neutrino masses $\Delta m^2_{23}$ is over an order of magnitude larger
than $\Delta m^2_{12}$, we will, for the time being, assume the latter
is zero. We thus have two neutrinos of the same mass $m_1=m_2$ and the
third one of a different mass $m_3$. We parametrise the masses in
terms of the sum of neutrino masses $\sum m_i$ and the fraction
$\alpha$ of the total mass in the third neutrino mass eigenstate, so that

\begin{equation}
  m_3=\alpha \sum m_i
\end{equation}

This particular parametrisation has been chosen, because it allows for
both extreme possibilities, namely that all of the mass is in the
third neutrino ($\alpha=1$) or that third neutrino is massless
($\alpha=0$). The value of $\alpha=1/3$ corresponds to the degenerate
case.

We use a modified version of the CAMB linear solver
\cite{2000ApJ...538..473L} that can evolve two families of neutrinos
of different masses separately \cite{2005astro.ph.11500D}. The
\texttt{accuracy\_boost} parameter was set to 2, which should result
in an accuracy around $0.1$\%. In addition, we used
\texttt{transfers\_high\_precision} option.  We have checked that the
results are the same regardless of whether a full hierarchy
integration is performed or a switch to series in velocity weight once
neutrinos become non-relativistic is used. Other parameters were set
to their nominal values for a $\Lambda$CDM universe. The CMB power
spectra discussed here were lensed using the algorithm discussed in
\cite{2005PhRvD..71j3010C}. 

In Figure \ref{fig:big} we plot the lensed CMB power spectrum, the
matter power spectrum and the projected gravitational potential power
spectrum for three models containing either three massless neutrinos
or three massive neutrinos with $\sum m_i=2$eV and $\alpha=1/3$ and
$\alpha=1$. The figures correspond to the standard flat $\Lambda$CDM
cosmology. The energy densities of baryonic and cold dark matter as
well as curvature were kept fixed so that hot dark matter component
grows at the expense of cosmological constant. The sum of neutrino
masses that large is not compatible with current observations and was
chosen for illustrative purposes only.  The effect on CMB is largely
determined by the change in the scale factor of the matter-redshift
equality ($a_{\rm eq}$), the change in which affects the CMB peak
positions (due to change in the sound horizon) and heights (via the
early integrated Sachs-Wolfe effect). However, the correction due to
non-degeneracy assumption is very small. The matter power spectrum and
the lensing potential power spectrum show very similar trends. If one
neutrino contains all the mass, it has a smaller free streaming length
and consequently damping does not extend to scales as large as in the
case where neutrinos have degenerate mass.  However, because the total
neutrino energy density is the same, the overall effect is the same
for $k\gg k_{\rm fs}$. The latter also implies that, contrary to what might
be naively expected, the precision measurements at very small scales
(such as those probed by Lyman alpha forest) will not be sensitive
probes of neutrino mass differences.

\begin{figure}
  \centering
  \includegraphics[angle=-90, width=\linewidth]{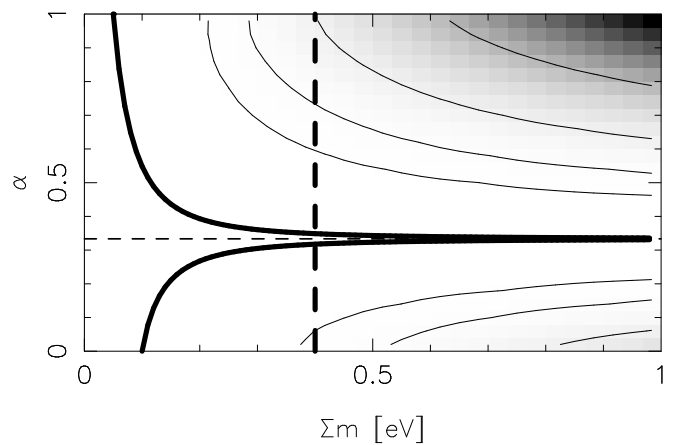}  
  \caption{This figure shows the change in the $\chi^2$ for a cosmic
  variance limited experiment (to $\ell=2000$) if one wrongly assumes
  degenerate neutrinos. The contours are at $\Delta \chi^2$ of 1, 5,
  25 and 125 (from $\alpha=1/3$ line outwards). 
 See
    text for discussion of other features on the plot.}
  \label{fig:cl}
\end{figure}

\section{Results}

How big are the discussed effects for realistic neutrino masses and experiments? To
answer this question we calculated power spectra for a grid of models
with $\sum m_i$ between 0 and 1eV and $\alpha$ between 0 and 1. In
each case the power spectra were calculated using both the
approximation that we have 3 degenerate neutrinos with a given $\sum
m_i$ and the correct distribution of neutrino masses, evolving the two
neutrino species separately.

For the CMB power spectrum, there exist a natural limit for the
accuracy with which the power spectrum can ever be measured. The so
called cosmic-variance is a result of a finite number of spherical
harmonic modes on the sky and is (for $\ell \gsim 50$) excellently
approximated by a Gaussian distribution with an error given by

\begin{equation}
  \sigma_{\rm CV} (C_\ell) = C_\ell \sqrt{\frac{2}{2\ell+1}}
\end{equation}

For each error in our parameter space we calculated the change in
$\chi^2$ induced by the correction stemming from non-degeneracy, i.e.
\begin{equation}
  \Delta \chi^2 = \sum_\ell \frac{(C_{\ell,{\rm  degen}}-C_\ell)^2}{\sigma_{\rm CV} (C_\ell)^2},
\end{equation}
where $C_{\ell,{\rm degen}}$ is the theoretical prediction if
degeneracy is assumed, for an experiment that is cosmic-variance
limited up to $\ell=2000$.  Such measurements could be performed with
a future experiment with a beam size of $4'$, temperature sensitivity
of $\Delta_{\rm T} \sim 1 \mu$K and polarisation sensitivity of
$\Delta_{\rm P} = \sqrt{2}\Delta_{\rm T} \sim 1.4 \mu$K
\cite{2002ApJ...574..566H}.  The results are plotted in the Figure
\ref{fig:cl}.  This figure is worthy some discussion. The horizontal
axis is the sum of the neutrino masses, while the vertical axis
correspond to the $\alpha$ parameter defined above. The dashed line at
$\alpha=1/3$ shows the degeneracy condition (where $\Delta \chi^2$ is
equal to zero by construction) and the vertical dashed line
corresponds to the tightest limit on the sum of neutrino mass found in
literature \citep{2005PhRvD..71j3515S}, that is $0.4$eV. A few recent
preprints find tighter limits using the latest cosmological data:
$0.3$eV in \cite{2006astro.ph..2155G} and $0.17$eV in
\cite{Seljak:2006bg}.  Solid thin lines are contours of constant
$\Delta \chi^2$ as stated in the caption.  The two solid thick lines
correspond to the value of $\alpha$ required to satisfy the $\Delta
m^2_{23}=2.5 \times 10^{-3}$eV$^2$ condition. The upper branch
corresponds to the normal hierarchy, while the lower branch
corresponds to the inverted hierarchy.

\begin{figure}
  \centering
  \includegraphics[angle=-90, width=\linewidth]{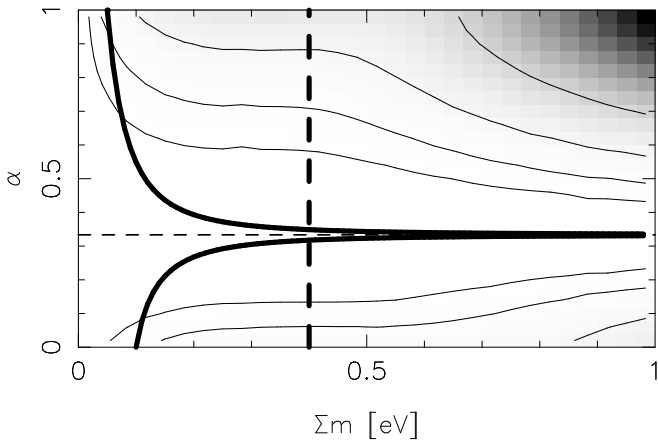}  
  \caption{Same as Figure \ref{fig:cl} but for the lensing potential
  reconstruction that is cosmic variance up to $\ell=1000$.}
  \label{fig:phic}

\end{figure}

This results shows the expected result that the magnitude of
correction due to the non-degeneracy is a function of both relative
mass difference and the total sum of the masses. If one takes
earth-based measurements of the mass square difference seriously, then
it seems that the standard degenerate neutrinos assumption is a good
one: the approximation is either saved by being a too small relative
effect at larger sum of masses or being a too small absolute effect
anyway, when the sum of masses is small. However, if one is not bound
by the small mass square difference (when considering exotic neutrino models,
for example), then the effect can be quite large and detectable with
high precision in the future CMB experiments.

A much more interesting picture emerges if one looks at the
reconstructed lensing potential. The reference experiment discussed
above could, ideally, reconstruct the lensing potential to a cosmic
variance limit up to $\ell=1000$ \cite{2002ApJ...574..566H}.  Figure
\ref{fig:phic} shows contours analogous to that of the Figure
\ref{fig:cl}, but for the lensing potential instead. One can see that
for $\sum m_i \lesssim 0.1$eV , the non-degeneracy corrections can
become important at high confidence. Our estimate is conservative
since it assumes that information on lensing potential is not
available beyond $\ell=1000$. A real experiment will still have
sensitivity to recover the lensing potential at $\ell>1000$, albeit
with a sub cosmic variance precision. We also note, that linear
transfer functions were used, but it is unlikely that non-linear
corrections would significantly destroy the sensitivity.

\begin{figure}
  \centering
  \includegraphics[angle=-90, width=\linewidth]{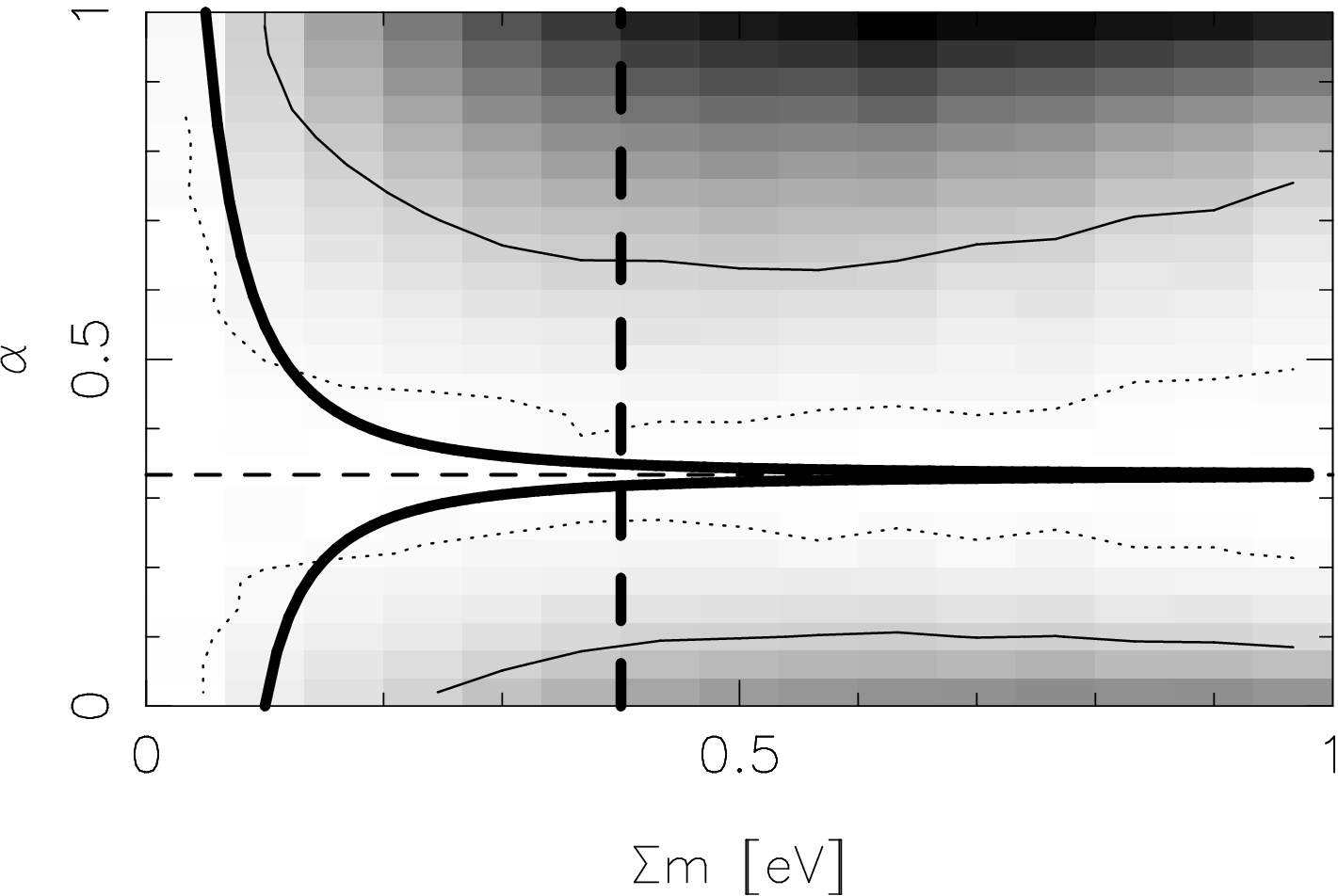}   \\
  \includegraphics[angle=-90, width=\linewidth]{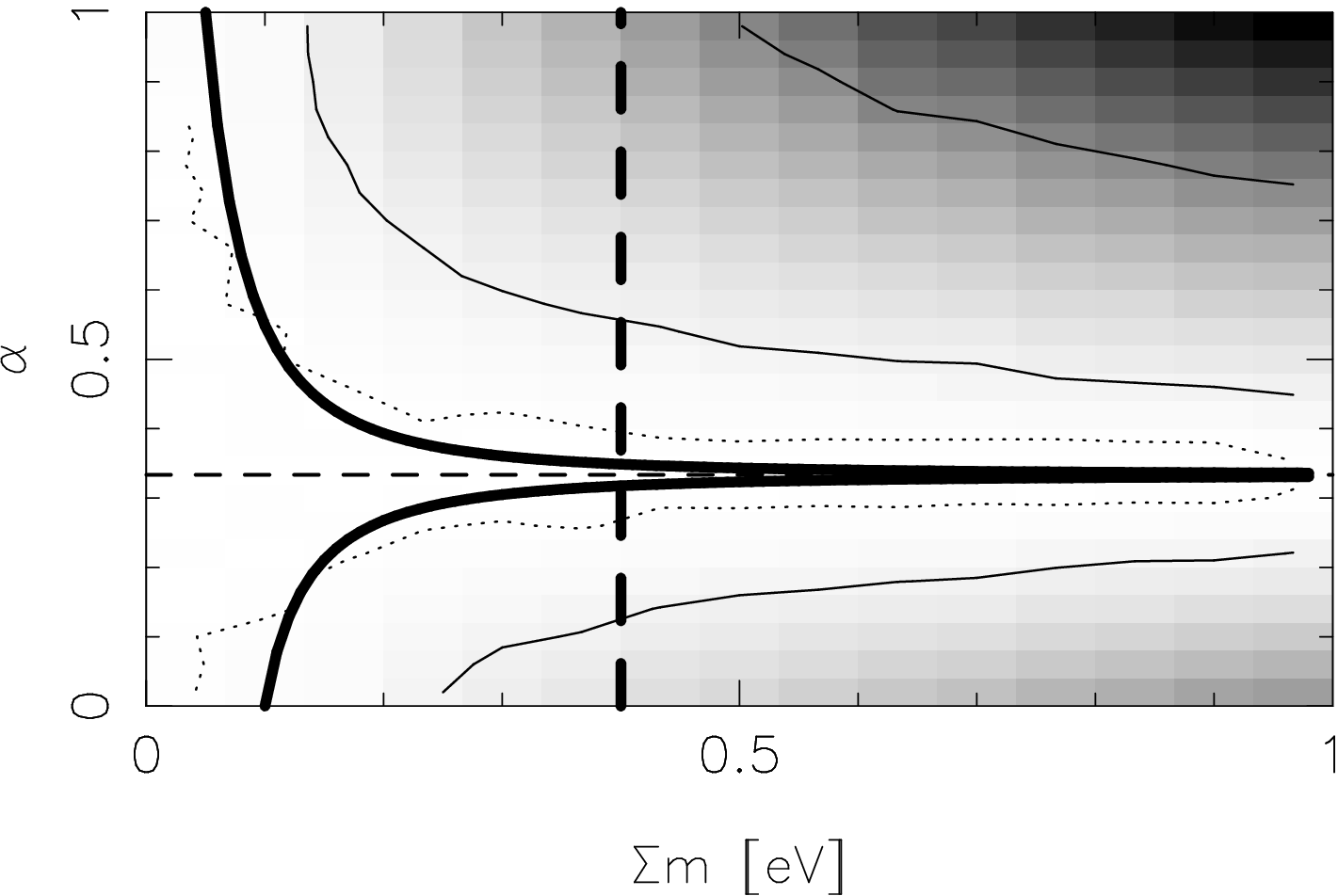}  \\
  \includegraphics[angle=-90, width=\linewidth]{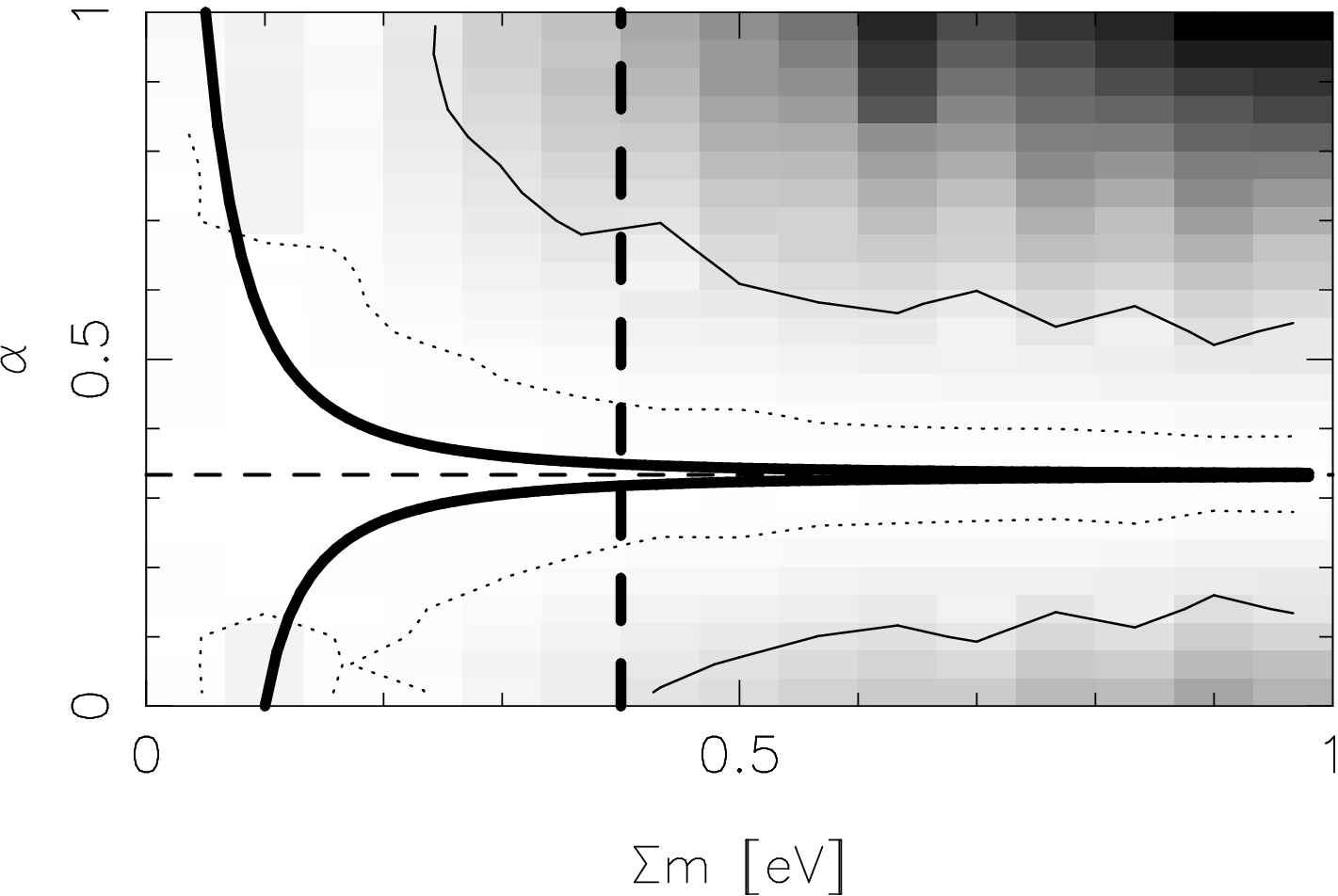}  \\
  \caption{This figure shows the relative change in the matter power
    spectrum slope at $k=0.005 h/$Mpc (top), $0.01 h/$Mpc (middle) and
    $0.1 h/$Mpc (bottom).  The thin solid and dotted lines correspond
    to 0.1\% (dotted), 1\% and 10\% difference and increase in value
    from the thin horizontal dashed line outwards.}
  \label{fig:pk}

\end{figure}

Next we turn to the matter power spectrum. The power spectrum is more
difficult to consider as it is not clear what an idealised experiment
can do in the presence of non-linear biasing and complications arising
from astrophysical considerations.  Therefore we calculate the
relative change of the slope of the linear power spectrum at three
nominal values of $k=0.005 h/$Mpc, $0.01 h/$Mpc and $0.1 h/$Mpc (where
$h$ is the reduced Hubble's constant). These are plotted in the Figure
\ref{fig:pk}. We see that the changes in the slope of the linear power
spectrum are of the order of a $\sim 0.1$\% and very likely
undetectable in the foreseeable future.  Again, we note, however, that
if the bound on the mass square difference is abandoned, the
non-degeneracy correction can be quite large and exceed the 10\% mark
in some cases.

\begin{table*}[htbp]
  \centering
  \begin{tabular}{ccccc}
    parameter & Normal Hierarchy & Inverted Hierarchy & Degenerate 1 & Degenerate 2\\
\hline
 & & & & \\
    $\sum m_i$/eV  & $0.055$ & $0.105$ &  $0.055$   & $0.105$  \\
    $\alpha$  & $0.95$ & $0.043$ &     $1/3$     & $1/3$\\
 & & & & \\
\hline \vspace*{0.1cm}
 & & & & \\
$(F_{\alpha\alpha})^{-1/2}$ & 0.1 & 0.1  & {} & {} \\ \vspace*{0.1cm}
$N_\sigma={\left(\alpha-\frac{1}{3}\right)}/{(F_{\alpha\alpha})^{-1/2}}$ & 4.9 & 2.9 & {} & {}\\ \vspace*{0.1cm}
$(F^{-1}_{\alpha\alpha})^{1/2}$ &  2.2 & 1.2 & {} & {} \\ \vspace*{0.1cm}
$N_\sigma={\left(\alpha-\frac{1}{3}\right)}/{(F^{-1}_{\alpha\alpha})^{1/2}}$ & 0.26 & 0.2 & {} & {}\\ 
 & & & & \\
\hline 
& & & & \\ \vspace*{0.1cm}
$(F_{\Sigma m\Sigma m})^{-1/2}$/eV & 0.003 & 0.003 & 0.003 & 0.003  \\ \vspace*{0.1cm}
$N_\sigma={\sum m}/{(F_{\Sigma m\Sigma m})^{-1/2}}$ & 18.8 & 36.6 & 18.8 & 36.6  \\ \vspace*{0.1cm}
$(F^{-1}_{\Sigma m\Sigma m})^{1/2}$/eV & 0.08 & 0.06 & 0.013 & 0.04 \\ \vspace*{0.1cm}
$N_\sigma={\sum m}/{(F^{-1}_{\Sigma m\Sigma m})^{1/2}}$ & 0.6 & 1.7 & 4.2  & 2.5\\
  \end{tabular}
  \caption{This table shows the results of fisher matrix analysis. See text for discussion.}
  \label{tab:fish}
\end{table*}

\section{Fisher matrix analysis}

In the above sections we have shown that in some parts of the
parameters space the effect of the difference in neutrino masses can
produce sizeable $\chi^2$ differences. This implies that neutrino mass
difference must be taken into account in order to avoid biases in the
data. However, it does not necessarily mean that neutrino mass
difference will be detectable due to possible degeneracies with other
parameters. In order to check for that effect we perform a Fisher
Matrix analysis. We use the following parametrisation of the model
\begin{equation}
  \theta_i = \left(\omega_{\rm b}, \omega_{\rm cdm}, h, \tau, n_{\rm s}, A,
  \sum m, \alpha\right),
\end{equation}
where parameters have their usual meaning in the cosmological context.
The Fisher matrix is given by
\begin{equation}
  F_{ij} = \sum \frac{\partial C^{X}_\ell}{\partial \theta_i}
  \frac{\partial C^{Y}_\ell}{\partial \theta_i} \left({\rm Cov}_\ell^{XY}\right)^{-1},
\end{equation}
where $XY$ is either TT, EE, TE (temperature, E polarisation and cross
power spectra) and LL (lensing potential cross-spectrum).  Since the exact
experimental parameters are not the focus of this work, we simply
assumed that the CMB TT power spectrum is known to cosmic variance for
$\ell<2000$, while TE, EE and lensing spectra are known to cosmic
variance for $\ell<1000$.

The interpretation of the Fisher matrix is straightforward: $(F_{ii})^{-1/2}$
is the expected $1-\sigma$ error on the measurement of the $i$-th
parameters, \emph{assuming all other parameters to be fixed}. The
value of $(F^{-1}_{ii})^{1/2}$  gives the expected $1-\sigma$ error on
the measurement of the $i$-th parameter taking into account possible
degeneracies with other parameters while the direction of these
degeneracies are given by the eigenvectors of $F$. A Gaussian nature
of the posterior is assumed throughout. 

We have performed the Fisher matrix analysis, using the nominal values
for most parameters, $\(\omega_{\rm b}, \omega_{\rm cdm}, h, \tau,
n_{\rm s}, A\)=\(0.02,0.12,0.7,0.11, 1.0,2.3\times10^{-9}\)$.  For the
values of the remaining two parameters we took two representative
points, one for the normal and one for the inverted hierarchy, which
also satisfy the atmospheric neutrinos constraint. We have
also performed the analysis by fixing $\alpha=1/3$ and excluding that
parameter from the analysis. The results are shown in the Table
\ref{tab:fish}.

There are a few interesting conclusions to be drawn. Firstly, we find
a reasonable agreement (given the crude approximation of experimental
performance) with \citep{Lesgourgues:2005yv} for the degenerate case,
where we find that the marginalised error on the sum of neutrino
masses will be of the order $0.013-0.04$ eV. Secondly, we note that the
degeneracies with other parameters completely destroy the detection of
the neutrino mass difference. From a modest few sigma detection, the
error increases several-fold. Thirdly, assuming
degeneracy severely decreases the accuracy with which the sum of
neutrino masses can be measured. Indeed, the analysis of the
eigenvectors of the Fisher matrix show that the two  strongest
degeneracies involve $\sum m$, $\alpha$ and $h$ in one eigenvector and
$\omega_{\rm dm}$, $\alpha$ and $h$ in the other. The latter is yet another face of
the degeneracy discussed in \citep{2005astro.ph.11500D}.

\section{Conclusions}

In this paper we have examined the importance of the degeneracy
assumption, which is often used in literature to constrain the sum of
masses of neutrinos from cosmology.  Since the combination of the
measurements of neutrino oscillations from the earth-based experiments
with the upper limit on neutrino masses from the cosmological
experiments imply that the neutrinos are non-degenerate at the level
of at least several percent, it is not obvious that this assumption is
a justified one.

By comparing model predictions for a model with two different neutrino
masses with that of an equivalent model with three degenerate
neutrinos we were able to show that the degeneracy assumption is
indeed valid for the constraints based on the CMB and galaxy power
spectra. In the case of CMB, this is true even for the cosmic variance
limited experiment, while the matter power spectra are unlikely to
reach the accuracy required to detect neutrino mass difference due to
various astrophysical constraints such as scale-dependent biasing.
Moreover, the size of the effect is at the level of numerical
precision of the present generation linear codes.

The best hope for detecting neutrino mass difference with cosmology
lies in the CMB lensing potential reconstruction.  In an idealised
future experiment such as that of \cite{2002ApJ...574..566H}, the
neutrino mass difference could bias the results of parameter
estimation for $\sum m_i \lesssim 0.1$eV, since it significantly
affects the $\chi^2$. However, by invoking Fisher matrix analysis we
have shown that it would be impossible to measure it due to degeneracy
with other parameters. These degeneracies could be broken by
constraining $\omega_{\rm dm}$ and $h$ in a independent manner. We
have also shown that once the degeneracy assumption is lifted, the
error on the measurement of the sum of neutrino masses is
significantly worsened; even for only two neutrino masses discussed
here.

If there is no information on the neutrino mass difference, as it is
usually the case in exotic scenarios, such as those involving extra
sterile neutrinos, then the corrections are potentially significant
for all cases. The slope of the matter power spectrum can change as
much as 10\% and the change could be detected with high significance
even for the CMB power spectrum if $\sum m_i \gsim 0.4$eV. However,
recent analysis have shown that the number and masses of neutrino
species are significantly constrained even with the present generation
cosmological data and thus we will not explore this further.

\section{Acknowledgements} Author acknowledges very useful discussions with
Uro\v{s} Seljak and Antony Lewis. He further thanks Antony Lewis for
``Englishication'' of the abstract. Additional thanks go to a very
thorough referee. Finally, the author thanks Oxford Astrophysics 
Kavli Institute for Particle Astrophysics and Cosmology for the hospitality
during which parts of this work were done.  AS is supported by the
Slovenian Research Agency.

\bibliography{cosmo,cosmo_preprints}

\end{document}